\documentclass{vgtc}                          

\ifpdf
  \pdfoutput=1\relax                   
  \usepackage{graphicx}                
  \DeclareGraphicsExtensions{.pdf,.png,.jpg,.jpeg} 
\else
  \usepackage{graphicx}                
  \DeclareGraphicsExtensions{.eps}     
\fi%

\graphicspath{{figures/}{pictures/}{images/}{./}} 
\usepackage{enumitem}
\usepackage{microtype}  
\usepackage{xcolor}
\PassOptionsToPackage{warn}{textcomp}  
\usepackage{textcomp}                  
\usepackage{mathptmx}                  
\usepackage{times}                     
\usepackage{cite}                      
\usepackage{tabu}                      
\usepackage{booktabs}      
\usepackage{tabularx} 
\usepackage{multirow}
\usepackage{balance}
\newcommand{\shri}[1]{\textcolor{black}{#1}}

\onlineid{0}

\vgtccategory{Research}

\vgtcinsertpkg

\title{Challenges and Opportunities of Teaching \\ Data Visualization Together with Data Science}

\author{Shri Harini Ramesh\thanks{e-mail: shrihariniramesh@cmail.caleton.ca}\\ %
        \scriptsize Carleton University %
\and Fateme Rajabiyazdi\thanks{e-mail:fatemerajabiyazdi@cunet.carleton.ca}\\ %
     \scriptsize Bruyere Research Institute\\ \scriptsize Carleton Unviersity}

\abstract{With the increasing amount of data globally, analyzing and visualizing data are becoming essential skills across various professions. It is important to equip university students with these essential data skills. To learn, design, and develop data visualization, students need knowledge of programming and data science topics. Many university programs lack dedicated data science courses for undergraduate students, making it important to introduce these concepts through integrated courses. However, combining data science and data visualization into one course can be challenging due to the time constraints and the heavy load of learning. In this paper, we discuss the development of teaching data science and data visualization together in one course and share the results of the post-course evaluation survey. From the survey's results, we identified four challenges, including difficulty in learning multiple tools and diverse data science topics, varying proficiency levels with tools and libraries, and selecting and cleaning datasets. We also distilled five opportunities for developing a successful data science and visualization course. These opportunities include clarifying the course structure, emphasizing visualization literacy early in the course, updating the course content according to student needs, using large real-world datasets, learning from industry professionals, and promoting collaboration among students.
} 
\CCScatlist{ 
Data Visualization, Data Science, Education, Challenges, Opportunities} %

\begin{document}



\maketitle

\section{Introduction}

In today's data-driven world, data visualization skills are highly valued and in demand across numerous industries, including finance, healthcare, marketing, technology, and academia~\cite{Sean2019DS, Eltweri2022Big}. It is important to equip university students with essential data visualization skills for their future careers~\cite{Mike2023Birth}. 

Students need a prior background in programming and data science to learn, design, and develop data visualization. However, students often lack strong programming and/or data science backgrounds.
Previous studies have explored various methods of teaching data visualization to students without a programming background and have offered solutions such as using GUI-based visualization tools, including Tableau ~\cite{Lo2019tutorial} or providing hands-on training for developing visualizations through projects~\cite{Dietrich2021project}, and activities~\cite{Knudsen2023activity, Sousa2021activity}. 

Moreover, some university programs lack a dedicated data science course, particularly at the undergraduate level, leaving students without knowledge of data science. 
Offering courses that combine data science and data visualization for these students can be instrumental.
Such courses need to focus on providing students with the skills of the entire data science pipeline, from data collection and cleaning to analysis, interpretation, as well as data visualization. 
However, combining data science and data visualization into one course can be challenging due to time constraints and the heavy load of learning both concepts. So, it is important to introduce and evaluate the effectiveness of such courses and share the findings with the visualization community to continuously improve the curriculum~\cite{Bach2024call}. 

Thus, we developed and evaluated the effectiveness of teaching a course that combines data science and data visualization. We conducted a survey of the enrolled students and analyzed the responses. 
From the survey results, we identified four challenges of teaching data visualization together with data science, including [CH1] difficulty in learning and using multiple tools in a single term, [CH2] heterogeneous proficiency with tools and libraries, [CH3] difficulty learning data science-related topics (i.e., artificial intelligence (AI),  machine learning (ML), and data visualization) in one course, and [CH4] challenges in selecting and cleaning a dataset. 
We also distilled five opportunities for enhancing future courses focused on teaching data visualization with data science, including [OP1] making the course structure clear and updating the content according to student needs, [OP2] introducing large real-world datasets for course assignments, [OP3] learning from industry professionals, [OP4] enhancing learning through collaboration, and [OP5] emphasizing visualization literacy and making interactive visualizations early in the course.
We believe that the challenges and opportunities identified in our study will provide a foundation for educators when teaching data visualization together with data science. 
\section{Course Background}
We introduced and developed a new course, ``Introduction to Data Science and Data Visualization,'' teaching both data science and data visualization to undergraduate students at an engineering department. This was a comprehensive 14-week course that included instructor-led lectures, class activities, guest lectures, and datathons (hands-on tutorial lab sessions). We held in-person lectures twice a week, each lasting 80 minutes; 20 minutes of each lecture was dedicated to class activities involving small tasks related to the lecture content, often using specific software or tools. We also invited guest speakers from academia, government, and industry—experts in their fields—to share their knowledge during lectures. Furthermore, we organized a total of 11 tutorial lab sessions, known as `Datathons,' held weekly. Each session lasted 180 minutes and took place in person. The prerequisites for this course were knowledge of statistics and programming fundamentals to ensure students had proficiency in software development in at least one language. 

\shri{
The learning outcomes for students in this course included:}
\begin{enumerate}[itemsep=0.1em]
    \item Explain different properties of data
    \item Collect, clean, and process data
    \item Analyze data using various techniques and tools
    \item Assess existing visualizations based on visualization theory and principles
    \item Design effective visual representations of data for exploration and communication
\end{enumerate}

We assessed students through five components: datathons, an individual project, a midterm, class activities, and guest lectures. Datathons comprised 50\% of the final grade. Students had to complete at least 10 datathons. The evaluation was based on participation, presentation of the results, and the ability to answer questions posed in the datathons.  The individual project was worth 25\% of the final grade. Students had to select their own dataset, analyze the dataset, develop a visualization, and find insights from the data. It had three deliverables: a 2-page preliminary report, a class demo, and a detailed final report. 
The midterm, contributing 20\% of the final grade, required students to achieve at least 50\% to pass. The midterm had seven questions totaling 100 marks, with an additional bonus question worth 10 marks. Class activities and participation in guest lectures made up the remaining 5\% of the final grade.

\subsection{Course Syllabus} 
The course had six modules: Data Collection and Storage, Data Cleaning and Integration, Data Visualization, Correlation and Regression, Clustering and Classification, and Text Analysis and Text Visualization. Each module was comprised of a lecture, a class activity, and a datathon.

\subsubsection{Module 1: Data Collection and Storage}
\textbf{Lectures:} Module 1 spanned two weeks. In the first week, we introduced the roles and responsibilities of a data scientist, the essential skills required for the profession, the interconnectedness of data science and data visualization, and an overview of the data science lifecycle—from data collection to visualization. In the second week, we focused on data storage concepts and Structured Query Language (SQL) fundamentals. We covered SQL basics (creating tables, managing data types, and writing data queries) and introduced a database engine, SQL Lite (SQLite). Additionally, we introduced modern computing infrastructures, Google Colab
, and cloud Tensor Processing Units (TPUs).

\textbf{Class Activities:} In the first week, we asked students to use tools such as the Application Programming Interface (API) Connector 
to extract data. 
In the second week, we asked students to write SQL queries for a given sample dataset, such as identifying products from specific countries and calculating prices.

\textbf{Datathons:} In the first week, we asked students to design survey questions using tools such as Google Forms and conduct the survey with their fellow students. To evaluate, we asked them to present the survey questions and responses, explain the reasoning behind the survey questions, discuss any challenges they faced during the survey conduction, and describe how they stored the survey data. In the second week, we asked students to write queries (e.g., create tables, delete entries) using SQLite and integrate SQLite with Python. They executed all these tasks using Python notebook in Google Colab. Then, to evaluate, we asked them to present their notebook and discuss the challenges they faced during this process.

\subsubsection{Module 2: Data Cleaning and Integration}

\textbf{Lectures:}  Module 2 spanned two weeks. In the first week, we had a guest panel of data scientists from the city of Ottawa who discussed the city's open data and the process of making data open to the public. We also lectured about the importance of data cleaning, sources of error in data, methods to prevent errors, and ways to identify errors. In the second week, we discussed data wrangling, data integration,  tools for data wrangling, and data integration.

\textbf{Class Activities:} In the first week, we conducted an activity where students had to measure and edit the distance between two string sequences. In the second week, we had activities for using OpenRefine to clean a dataset and using Tableau to integrate and join datasets.

\textbf{Datathons:} In the first week, we asked students to work on cleaning data using a Python notebook in Google Colab. We provided a dataset and asked them to identify missing values and handle them using various data-filling methods. We also asked them to deal with duplicates and find outliers using visualization and methods such as z-score and interquartile range. To evaluate, we asked them to present the identified outliers in the data through visualization and discuss the potential impact of these outliers on the data analysis. In the second week, we asked students to connect various data sources in Tableau and create visualizations to conduct a preliminary analysis of the data. Students later presented their findings and conclusions based on the insights extracted from the visualizations.

\subsubsection{Module 3: Data Visualization}

\textbf{Lectures:} Module 3 spanned two weeks. In the first week, we discussed the importance of data visualization, effective and ineffective visual representations, and designing data visualizations. In the second week, we covered cognitive bias, visualization mirages, equity awareness in visualization, and an introduction to statistics.

\textbf{Class Activities:} In the first week, we conducted an activity in which students visualized two quantities to learn different methods for visualizing data~\cite{rockcontent45ways}. In the second week, we used a public dataset available in Tableau and asked the students to visualize it to analyze the data and identify any mirages in the visualization.

\textbf{Datathons:} In the first week, we asked students to create a visual storyboard in Tableau using the provided dataset to gain insights and address questions presented in the datathon about the dataset.
In the second week, we provided a different dataset and asked students to create visualizations using Power BI. They had to create visualizations for the questions posted in the datathon and an additional analysis question of their own that presented interesting insights about the data. For both of the week's evaluations, we asked them to present their findings and conclusions based on the insights extracted from the visualizations.

\subsubsection{Module 4: Correlation and Regression}

\textbf{Lectures:} Module 4 spanned three weeks. In the first week, we covered topics including the relationship between statistics and data science, P-hacking, power analysis, and exploratory data analysis. In the second week, we focused on correlation, regression analysis, and regression models. In the third week, we covered dimensionality reduction, principal component analysis (PCA), and T-distributed stochastic neighbor embedding (T-SNE).

\textbf{Class Activities:} In the first week, we asked students to critique a visualization they found online. In the second week, we asked students to perform regression analysis and visualize the results using Tableau Pulse, an AI-driven, intuitive tool for exploring datasets and automating insights discovery.
In the third week, we had two activities. For the first activity, we provided students with a dataset and asked them to conduct regression models and visualize the results in Rstudio, For the second activity, we played a game with color names, teaching the Stroop effect~\cite{stroop1935studies}.

\textbf{Datathons:} In the first week, we asked students to explore frequency distributions, visualize categorical data, and answer statistical questions (e.g., mean, standard deviation, normal distributions) using Python notebooks in Google Colab. In the second week, students implemented linear and polynomial regression and performed exploratory data analysis.
In the third week, we asked students to implement multiple linear regression and create interactive visualizations for PCA representation. For all three weeks, we evaluated students by the quality of insights they shared when answering the data analysis questions for the provided dataset.

\subsubsection{Module 5: Clustering and Classification}

\textbf{Lectures:} Module 5 spanned two weeks. In the first week, we discussed clustering, its applications, and implementation. In the second week, we covered topics including classification, clustering, and their differences.
We also had a guest speaker, a technologist aiming to make technology understandable for the next generation. They spoke about the AI hype cycle in the industry and the ethical considerations that are often overlooked in the rush to be the biggest, the best, and the first.

\textbf{Class Activities:} In the first week, we invited two guest speakers: a chief executive officer from a data analytics company and a representative from Career Services. The motivation behind these guest lectures was to enhance student's understanding of various data scientist roles, their day-to-day responsibilities, available jobs and roles in the industry and academia for data science, and the required skills. Students were encouraged to participate in the discussions and interact with the speakers. In the second week, we introduced students to a visualization tool called EduClust~\cite{Educlust2019}, a visual education platform for clustering algorithms. We asked students to experiment with different clustering algorithms and parameters using EduClust tool.

\textbf{Datathons:} In the first week, students implemented k-means clustering k-nearest neighbors, addressed overfitting and underfitting, and used decision trees, random forests, and naive Bayes classifiers using Python notebooks in Google Colab. To evaluate, we asked several questions, which students needed to answer by analyzing the provided dataset and then sharing the insights in class. In the second week, we didn't have a datathon because it was a holiday.

\subsubsection{Module 6: Text Analysis and Text Visualization}

\textbf{Lectures:} Module 6 spanned one week. we introduced text data, text data processing, named entity recognition, document-level metrics, term frequency-inverse document frequency, and text visualizations.

\textbf{Class Activities:} In this week, we conducted three activities. For the first activity, we asked students to choose one text visualization from online resources and critique it for the class. For the second activity, we asked students to upload a document into Voyant
~\footnote{ \href{https://github.com/voyanttools/Voyant}{https://github.com/voyanttools/Voyant}}
, a text analysis tool, visualize the text data using this tool and share their findings with the class. Then, for the third activity, we asked students to work with Julia
to visualize text data.

\textbf{Datathons:} In this week, we students worked with text data in Tableau, where they cleaned, processed, and visualized the frequency and sentiment in the text data. To evaluate, we asked them to share the insights gained by visualizing the text data.

\section{Survey Study}
We conducted an anonymous survey with students enrolled in the 14-week long ``Introduction to Data Science and Data Visualization'' course in Winter 2024. Our aim was to assess the effectiveness of the course, especially in equipping undergraduate students with both data science and data visualization skills in a single course. 
We used Qualtrics to develop the survey, which included a mix of twelve Likert agreement scales ranging from strongly disagree to strongly agree, seven open-ended questions, and five multiple-answer questions. We began with demographic questions, including gender, age, academic program, and year of study. We then asked questions about each part of the course, including the datathons, the individual project, the tools learned, and the overall course structure. 
During the last lecture of the course, we shared our study recruitment poster with the students containing details about the study and a QR code linking to the survey. Interested students could scan the QR code, which directed them to the survey after they signed a consent form. Data collection took place in April 2024, and participants received a \$10 CAD e-gift card as compensation for their participation in the survey. Once we collected the data, we cleaned and performed statistical analysis for the Likert scale and multiple-choice questions in Qualtrics. We also used open coding~\cite{goodman1961snowball} and affinity diagram technique~\cite{hartson2012ux} to analyze the open-ended questions.

\section{Results}

Out of 26 students enrolled in the class, 21 participated in the survey. 
The participants' demographics breakdown is provided in Table \ref{tab:participants}. \shri{The students had basic knowledge in fundamental programming, object-oriented programming, programming with Python and Java, algorithms, data structures, and statistical analysis from their previous years' coursework in their undergraduate studies.}

\begin{table}[h]
\centering 
  \caption{Demographic breakdown of survey participants.}
  \label{tab:participants}
  \scriptsize
  \centering
  \begin{tabular}{%
    p{0.08\textwidth}%
    p{0.2\textwidth}%
    p{0.08\textwidth}%
  }
    \toprule
    \textbf{Category} & \textbf{Details} & \textbf{Responses} \\
    \midrule
    \multirow{2}{*}{Age} & 18-24 years & 20 \\
     & 25-34 years & 1 \\
    \midrule
    \multirow{2}{*}{Gender} & Male & 15 \\
     & Female & 6 \\
    \midrule
    \multirow{3}{*}{Program} & Software Engineering & 15 \\ &Computer Systems Engineering & 5 \\
     & Electrical Engineering & 1 \\ 
    \midrule
    \multirow{2}{*}{Year of Study} & 4th Year & 17 \\
     & 5th Year & 4 \\
    \bottomrule
  \end{tabular}
\end{table}

\begin{table*}[ht]
\centering
\caption{Response of the students to the Likert agreement scale questions.}
\label{tab:feedback}
\small 
\begin{tabular}{%
    p{0.13\textwidth}%
    p{0.4\textwidth}%
    p{0.06\textwidth}%
    p{0.06\textwidth}%
    p{0.06\textwidth}%
    p{0.06\textwidth}%
    p{0.06\textwidth}%
  }
\toprule
\textbf{Topic}  & \textbf{Statement} & \textbf{Strongly Disagree} & \textbf{Disagree} & \textbf{Neutral} & \textbf{Agree} & \textbf{Strongly Agree} \\
\midrule
(i) Course structure &  Overall organization and structure of the course were clear and easy to follow & 0\% & 0\% & 10\% & 62\% & 29\% \\
\midrule
(ii) Datathons &  Lectures and datathons complement each other & 0\% & 5\% & 24\% & 57\% & 14\% \\
 &  Datathons encouraged collaboration and discussion among students & 0\% & 0\% & 14\% & 43\% & 43\% \\
 &  Group participation in datathons enhanced my learning experience and made problem-solving easier & 5\% & 10\% & 10\% & 52\% & 24\% \\
\midrule
(iii) Individual project  & The lectures and datathons helped me to develop the skills necessary to undertake individual projects independently & 0\% & 0\% & 29\% & 48\% & 24\% \\
 &  The individual project provided a valuable practical experience that reinforced my understanding of course concepts & 0\% & 0\% & 10\% & 48\% & 43\% \\
 & The individual project allowed me to explore topics of personal interest within the scope of the course & 0\% & 0\% & 0\% & 52\% & 48\% \\
\midrule
(iv) Skills and tools & The course equipped me with the skills and knowledge required to analyze various datasets & 0\% & 0\% & 5\% & 71\% & 24\% \\
\midrule
(v) Career aspirations & This course inspired me to pursue a career in data science and data visualization& 0\% & 14\% & 48\% & 33\% & 5\% \\
\midrule
(vi) Student confidence
 & I am confident in my ability to use Python for analysis and visualization in the future & 0\% & 0\% & 14\% & 24\% & 71\% \\
 & I am confident in my ability to use Tableau/Power BI for analysis and visualization in the future & 0\% & 0\% & 29\% & 62\% & 10\% \\
\bottomrule
\end{tabular}
\end{table*}

We present the results as follows: (i) Course structure and content, (ii) Datathons, (iii) Individual project, (iv) Skills and tools, (v) Career aspirations, and (vi) Student confidence. A summary of the results for the Likert scale questions is provided in Table \ref{tab:feedback}. 

\begin{table}[h]
\centering
\caption{Responses to the multiple choice question ``Which specific modules of the course do you feel most confident in after completing the course?'' with choices corresponding to the six course modules.}
\label{tab:student_confidence}
\scriptsize
\centering
\begin{tabular}{%
  p{0.25\textwidth}%
  p{0.15\textwidth}%
}
  \toprule
    \textbf{Modules} & \textbf{Responses} \\
    \midrule
    Data collection and storage & 43\% \\
    Data cleaning and integration & 71\% \\
    Data visualization & 90\% \\
    Correlation and regression & 19\% \\
    Clustering and classification & 19\% \\
    Text analysis and text visualization & 5\% \\
  \bottomrule
\end{tabular}
\end{table}

\textbf{ (i) Course structure and content: }Teaching data science and data visualization in one course required us to cover many topics within a term. Thus, one of our aims when developing the course was to provide a clear and easy-to-follow structure to ensure students could keep up with the topics without feeling overwhelmed. This was reflected in the feedback, with the majority of students, 62\% agreed, 29\% strongly agreed with the statement ``Overall organization and structure of the course were clear and easy to follow''. 

\textbf{(ii) Datathons: }Datathons were developed to complement lectures and provide hands-on experience with the concepts and tools learned in class. When asked if ``Lectures and datathons complement each other'', 57\% of students agreed, and 14\% strongly agreed. However, 5\% disagreed, indicating that there is room for improvement in integrating these components. 
Furthermore, datathons aimed to foster collaboration among students. In our survey, more than 80\% of the students (43\% agreed and 43\% strongly agreed) reported that ``Datathons encouraged collaboration and discussion among students''. 
In another survey question, 15\% of the students disagreed with the statement, ``Group participation in datathons enhanced the learning experience and made problem-solving easier.'' This indicates that although most students found group work beneficial, some students encountered challenges with collaboration. This could be because, in datathons, a few students took control of creating and presenting visualizations to ensure high grades, which limited opportunities for others to showcase their skills.

\textbf{(iii) Individual project: }The individual project component was developed for students to independently apply the skills and knowledge acquired throughout the course and develop a visualization system for their chosen dataset. When asked to respond to the statement, ``The lectures and datathons helped me to develop the skills necessary to undertake individual projects independently,'' 72\% of the students affirmed that the components developed the requisite skills for dataset analysis and visualization creation. More than 90\% of the students (48\% agreed and 43\% strongly agreed) reported that the ``The individual project provided a valuable practical experience that reinforced the understanding of course concepts.''. Furthermore, all students agreed to the statement, ``The individual project allowed me to explore topics of personal interest within the scope of the course.'' Some students even created their own datasets by collecting data from various online sources and literature, demonstrating their ability to independently gather and manage data.

\textbf{(iv) Skills and tools: }The course aimed to provide students with essential data science and data visualization skills and tools to analyze and visualize data. In the survey, 95\% of the students agreed with the statement, ``The course equipped me with the skills and knowledge required to analyze various datasets.''


\textbf{(v) Career aspirations: }The course was developed for fourth and fifth-year students who were close to graduating within 1-2 terms and pursuing a career. When asked to respond to the statement, `` This course inspired me to pursue a career in data science and data visualization''. Responses were mixed, with 33\% agreeing and 5\% strongly agreeing that the course inspired them to pursue a career in these fields, while 14\% disagreed, and almost half of the students remained neutral. While the course successfully motivated a substantial portion of participants, it’s understandable that not all students would envision these fields as their future careers after a single course. Career decisions often evolve over time and are influenced by various factors beyond academic exposure. Encouraging even a fraction of students to consider these career paths can be seen as a positive outcome. 

\textbf{(vi) Student confidence: }Throughout the course, students were taught various tools for analysis and visualization, including Python, Tableau, and PowerBI. When asked to respond to the statements, ``I am confident in my ability to use Python for analysis and visualization in the future'' and ``I am confident in my ability to use Tableau/PowerBI for analysis and visualization in the future,'' students expressed greater confidence in their ability to use Python (96\%) compared to Tableau or PowerBI (72\%). This could also be due to the fact that a few students had prior knowledge and hands-on experience with Python, whereas Tableau and PowerBI were new to most of them.
In addition to these questions, we also asked a multiple-choice question, `` Which specific modules of the course do you feel most confident in after completing the course?'' (Table ~\ref{tab:student_confidence}). Students expressed that they are most confident in data visualization, with 90\% feeling assured. However, confidence was low (19\%) in correlation and regression, as well as clustering and classification modules. The lowest confidence was in text analysis and visualization, with just 5\% feeling confident. The lower confidence in these topics can be attributed to the complex statistical and machine-learning concepts involved. Moreover, covering a broad range of topics within a term by combining data science with data visualization may have impacted the depth of coverage and practical experience gained in these advanced areas. 

\section{Discussion}
We analyzed the results of our survey and identified 4 challenges and 5 opportunities in teaching data visualization together with data science in one course (Table \ref{tab:challenges_opportunities}).

\subsection{Challenges}
\textbf{CH1: Difficulty in learning and using multiple tools in a single term- }To provide students with essential data science and data visualization skills, it was necessary to teach a wide range of tools. The objective of this introductory course was to give students a basic knowledge of data visualization tools to inspire them to explore further if they were interested. However, due to the short timeframe, students found it challenging to learn the tools and effectively utilize them. For example, one student mentioned, \textit{``I think personally grasping how to use the tableau was difficult. Lots of usefulness in the program, but maybe it was only for me, but trying to take the visualization ideas in my head to tableau was hard''} (P21). Students also encountered difficulties when using these tools for their individual projects. As one student stated, \textit{``Challenges with the individual project involved using Tableau/PowerBI effectively to visualize the data''} (P10). While the goal of the project was to give them an overview of the entire data analysis pipeline, from data collection to visualization, it was not meant to evaluate their ability to create advanced visualizations. However, students became interested in delving deeper into the tools as they worked on their projects, wanting to create more advanced visualizations. This posed a challenge for them within the given time, \textit{``[challenges I encountered while working on my individual project was] mainly figuring out how to do more complex operations within tableau to create the visualizations that I envisioned''} (P14). These experiences illustrate the tension between the introductory nature of the course and the students' desires to explore further. Echoing the results of previous research~\cite{Lo2019tutorial}, our survey results also highlight the struggle in learning and using multiple data analysis and data visualization tools in one term. Grouping students into different groups where they can choose their tools and learn about them could be a solution. This approach has also been offered in previous studies~\cite{Hedayati2023D3}.

\begin{table}[t]
\centering
\caption{Challenges and opportunities in teaching data visualization together with data science.}
\label{tab:challenges_opportunities}
\scriptsize
\begin{tabular}{p{0.03\textwidth}p{0.4\textwidth}}
\toprule
\textbf{Tag} & \textbf{Challenges and Opportunities} \\
\midrule
CH1 & Difficulty in learning and using multiple tools in a single term \\
CH2 & Heterogeneous proficiency with tools and libraries   \\
CH3 & Difficulty learning data science-related topics (i.e., AI,  ML, and data visualization) in one course\\
CH4 & Challenges in selecting and cleaning a dataset \\
\midrule
OP1 & Making the course structure clear and updating the content according to student needs \\
OP2 & Introducing large real-world datasets for course assignments\\
OP3 & Learning from industry professionals \\
OP4 & Enhancing learning through collaboration \\
OP5 & Emphasizing visualization literacy and making interactive visualizations early in the course \\
\bottomrule
\end{tabular}
\end{table}

\textbf{CH2: Heterogeneous proficiency with tools and libraries- }When developing the course, one of our goals was to ensure that every student could learn the various tools and libraries, regardless of their prior knowledge. 
The datathons were structured to guide students through tasks that ranged from operations, such as uploading data to Google Colab, to more complex processes involving statistical analysis and visualization. For some students, the tasks seemed too simple and basic, as they were already familiar with these operations. For instance, they commented that \textit{``Many datathons were fairly easy''} (P18), and \textit{``Some of the datathons seemed repetitive''} (P4). On the other hand, for students who were new to Python and data visualization, the same tasks posed a challenge. They felt overwhelmed at times and suggested: \textit{``…making the datathons slightly shorter would make it easier''} (P10). Despite receiving detailed explanations for advanced concepts, some students felt that they needed more time to understand them effectively. This reflects the challenge of varying levels of complexity each student faced with the same tools and libraries. This challenge echoes previous research, which emphasized that methods effective for one group of learners may not be as effective for another~\cite{Bach2024call} and that students often show varying levels of proficiency with visualization tools~\cite{Seo2024blind}. Offering free-form courses where students can choose their projects, libraries, and tools depending on their proficiency could be one solution~\cite{Beck2022critical}.

\textbf{CH3: Difficulty learning data science-related topics (i.e., AI,  ML, and data visualization) in one course- }Our course covered a wide array of topics in data science, ranging from foundational statistics such as mean, median, and mode to advanced topics such as ML and AI, along with data visualization. 
Many students struggled with the advanced topics. e.g., \textit{``Some of the machine learning stuff was a bit challenging''} (P9), reflecting the difficulty they encountered when delving into ML algorithms and techniques. Moreover, certain ML and AI topics require knowledge of statistics. Although students typically learn foundational courses such as statistics and mathematics in their first year, they often lose familiarity over time. Similarly, previous studies showed students from non-computer science backgrounds without a strong mathematical foundation found some data science topics difficult to understand~\cite{Velaj2023DS}. In our survey, one student mentioned, \textit{``Some of the correlation and regression concepts were initially hard to grasp without a refresh on basic statistics''} (P16). Students expressed interest in refreshing their understanding of some statistics and mathematical concepts within the course. Even though basic statistics were covered in the course, it’s not feasible to cover all the foundational math and statistics in the given timeframe. Doing so would have left even less time for other topics. This reflects the challenge of teaching and learning a wide range of data science topics, especially the advanced ones, along with data visualization. Data visualization itself is a complex and advanced topic, and learning it with other essential data science topics could add extra difficulties. 
While it might be impractical to require students to take a data science course before enrolling in a data visualization course, dedicating a few sessions at the beginning of the term to assess students' proficiency in foundational topics and refresh those topics as necessary could support students in learning advanced topics throughout the term.

\textbf{CH4: Challenges in selecting and cleaning a dataset- }An important step for learning how to design and develop a data visualization is to create a data visualization from scratch, find a dataset, or collect data and clean the data. 
Despite covering topics such as data selection and cleaning in the course, students encountered challenges in finding appropriate datasets for their individual projects. The abundance of available datasets often leads to choice paralysis, where students find it difficult to pinpoint one that aligns with their interests. As one student noted, \textit{``finding what dataset interested/that I wanted to visualize was hard. With lots of datasets available, it was hard for me to differentiate what would be best for me to select''} (P21). Moreover, even after selecting a dataset of interest, students frequently faced issues concerning its quality and size. Some datasets were too small to derive meaningful analysis from, e.g., \textit{``The dataset I chose was not large enough to extract more than 6 or so insights, I was not able to overcome this issue’’} (P6). Additionally, some students found their chosen datasets to be excessively complex and filled with inaccuracies, requiring substantial time for thorough cleaning and organization before they could commence their analysis and visualization tasks. One student expressed their challenge, \textit{``How to handle mistakes in real datasets?''} (P8). Despite their prior experience with partially cleaned datasets in datathons and class activities, students encountered difficulties when dealing with messy real-world data they collected. This aligns with previous research emphasizing the complexities of data acquisition and preparation in project-based data visualization education~\cite{Dietrich2021project}. This underscores the challenge of working with curated/partially cleaned datasets in tutorials and handling the intricacies of raw data sourced from real-world scenarios.
It is important for the community to curate a list of recently collected datasets that could serve as practice for students in data science and data visualization courses. These datasets should have an appropriate level of complexity for learning purposes, not to overwhelm students, and be exciting at the same time.

\subsection{Opportunities}
\textbf{OP1: Making the course structure clear and updating the content according to student needs- }To equip students with essential data science and data visualization skills, it is crucial to cover a broad range of topics in both fields. However, the need to complete the course within one term introduced time constraints. Therefore, efforts were taken to provide a clear and easy-to-follow structure, ensuring students could keep up with the topics without feeling overwhelmed. Students appreciated this course structure, noting, \textit{``The course was well structured and covered a wide range of relevant topics related to the field''} (P16). Another student expressed, \textit{``It was honestly the most enjoyable course I’ve taken in my university career. For the first time, I didn’t dread having to attend classes''} (P14). This highlights the importance of making the course structure clear to the students early in the course to set the right expectations for students. During the term, we got informal feedback from students on different aspects of the course and made changes to the ways we presented the content as necessary. One student highlighted, \textit{``I really appreciate feedback on the go and updating the course and datathons based on what students would like to learn''} (P19). This underscores the importance of flexibility and adapting the content of the course based on student's skills and needs. Future courses should prioritize making the course structure clear to the students while also being adaptable to change the content to student feedback, as highlighted in our survey with students and also in previous research in visualization education~\cite{Syeda2023coursecontent, Bach2024call} With students coming from various prior knowledge levels, it’s essential to gather feedback continuously and update the course content based on students' needs. 

\textbf{OP2: Introducing large real-world datasets for course assignments- }Students highly valued the practical experience gained from datathons, as they provided real-world applications beyond theoretical knowledge. One student noted, \textit{``Datathons proved a 'real-world' use case beyond the theoretical. We had the chance to start from a messy dataset and develop it into a clean visualization''} (P11). While datathons aimed to simulate real-world scenarios, they often utilized small datasets to help them with the learning curve. This approach was made to ensure that students new to data science and data visualization could learn foundational concepts without being overwhelmed by the complexity of handling vast amounts of data. However, students expressed a growing interest in working with larger datasets to gain more real-world hands-on experience. One student suggested, \textit{``Maybe introduce larger datasets ... Maybe also explore some pipeline aspects (e.g., pulling data from AWS, Azure)''} (P14). Future data science and visualization courses should consider utilizing large real-world datasets, especially \shri{with students who have a background in engineering and programming.} Integrating larger datasets into future courses not only better prepares students for real-world scenarios but also familiarizes them with the complexities involved, such as scalability, readability, and the substantial computing power often required~\cite{Strahler2020dataset, Burlinson2016bridges}. \shri{For students from non-engineering or programming backgrounds, caution should be exercised when introducing large datasets.}

\textbf{OP3:  Learning from industry professionals- }Guest lectures were incorporated into the course to provide students with diverse perspectives from industry, government, and academia. In regular lectures, students gain knowledge of specific topics and theoretical foundations for data science and visualization. However, guest lectures bring an added dimension to the concepts they've learned in class~\cite{krogstie2018guest}. \shri{Guest lectures from industry professionals are particularly beneficial in data science and data visualization courses as these skills are highly demanded in various industries. Firsthand insights into industry practices can bridge the gap between theory and practice, highlighting the evolving demands and applications in the field.} One student highlighted, \textit{``I also really enjoyed the guest lectures, having people in the industry talk to us from both the private and city of Ottawa was cool''} (P21). While the guest lectures were generally well-received, some feedback indicated variability in their effectiveness e.g., \textit{``Some guest lecturers were better than others. For example, the individuals from [data analytics company] were very good, but the one about making presentations was somewhat repetitive to things we have already been taught in previous courses'' }(P18). The feedback highlights the importance of guest lectures and students' interest in learning directly from industry experiences rather than solely focusing on acquiring specific skills related to student success. This aligns with previous research indicating that students have shown interest in introducing industry guest lectures into the mass communication curriculum~\cite{Patrick2017guest}.  Thus, future courses should prioritize guest lectures from professionals who are working with real-world data, perhaps in industrial settings. \shri{We also acknowledge that it could be challenging to bring industry people into the classroom, and alternative methods, such as virtual guest lectures or industry collaboration site visits, can be explored.}

\textbf{OP4: Enhancing learning through collaboration- }Collaboration among students is helpful in exchanging ideas and approaches, enhancing their learning experience~\cite{Friedman2021LeveragingPR}. In this course, all the datathons and many class activities were conducted in groups to emphasize the importance of collaboration and to make problem-solving easier. One student appreciated this aspect, saying, \textit{``Working in groups allows us to compare ideas and approaches; which visualizations work and which do not''} (P11). Collaborative learning in datathons helped students peer review each other's work, provide comments, and inspire one another. They also supported each other during their individual projects.
When one student was stuck, another could offer assistance, fostering an environment of collaborative learning.  For instance, one student shared, \textit{``…find[ing] good insights for the project [was challenging]… I overcame it by getting my peers’ opinions on graphs I formed and explained''}(P5). However, some students expressed a desire to work alone at times while still having access to peer support. One student remarked, \textit{``A lot of the datathons were more beneficial to complete individually (the group aspect was beneficial to ask questions if we got stuck, but we always each did it individually)'' }(P13). 
Although research in data visualization education supports and encourages group work for active learning and exposure to diverse student work~\cite{Burch2020group, Beasley2021peer}, feedback from our students highlights the importance of providing both group work and individual tasks. For a successful data science and data visualization project, a variety of skills, such as design, development, and engineering, are necessary, making collaboration useful. However, it can be challenging to divide tasks properly and fairly and to ensure all group members learn everything. Future courses should incorporate both collaborative activities and opportunities for students to work independently. This approach will provide a comprehensive understanding, allowing students to benefit from peer insights while also developing their individual skills and confidence in tackling data science and data visualization challenges.

\textbf{OP5: Emphasizing visualization literacy and making interactive visualizations early in the course- }It’s becoming increasingly important to teach visualization literacy, which involves the ability to read, interpret, and create data visualizations~\cite{Brner2019DataVL}. It goes beyond simply understanding data representations and includes the capacity to critically evaluate, interact effectively, and thoughtfully design and develop data visualizations~\cite{Boy2014Vizlit}. Although the course was structured to cover data visualization topics in the third and fourth weeks, students expressed a strong desire to learn these skills earlier. One student emphasized, \textit{``Have all the ‘How to make good data visualization’ classes at the beginning.''} This feedback highlights the need for a knowledge of visualization literacy from the start. During the course, students showed an interest in learning about good practices in creating `right' data visualizations. Due to student interest, we added a lecture on ``Good practices in creating 'right' data visualizations'' midway through the course, during the datathon, whenever time allowed. 
Additionally, students expressed a desire to learn how to create more aesthetically pleasing and dynamic visuals, with one student noting, \textit{``Teach us how to make more pretty and dynamic visuals''} (P1). These insights highlight the strong interest students have in learning how to make interactive data visualizations. Previous research has also emphasized the importance of supporting the development of skills and knowledge in interactive exploration~\cite{Bach2024call}. Future courses should aim to cover these topics more thoroughly and emphasize them early on, ensuring that students are well-equipped to create effective visualizations from the start.

\section{Conclusion}
When teaching data visualization to university-level students who lack prior knowledge of data science, it is essential to incorporate data science concepts into the visualization curriculum. 
Thus, our aim was to develop and evaluate the effectiveness of teaching a course that combines data science and data visualization.
We conducted a survey with the students enrolled in the course. In this paper, we described the development of a course that combines data science and data visualization and shared the survey results. Based on our results, we identified four challenges in teaching data visualization together with data science. we also distilled five opportunities for enhancing future courses focused on teaching data visualization together with data science.
We hope that the challenges and opportunities highlighted in our study will provide valuable insights for educators and researchers in the visualization community. \shri{However, we acknowledge that these challenges and opportunities are specific to the context of teaching data visualization together with data science in a single course and may not be generalizable to other courses.}


\section*{Acknowledgements}
We sincerely appreciate the valuable input of all survey participants. Additionally, we acknowledge the generous support provided by Carleton University to conduct this research.

\bibliographystyle{abbrv-doi}

\balance
\bibliography{References}
\end{document}